\documentclass[12pt]{article}
\usepackage{t1enc}
\usepackage[latin1]{inputenc}
\usepackage[english]{babel}
\usepackage{graphicx,epsfig}
\usepackage{a4}
\usepackage{fancyhdr}
\input epsf
\begin{document}

\addtocounter{footnote}{1}

\title{Stable branches of a solution for a fermion on domain wall}

\addtocounter{footnote}{0}

\author{V. A. Gani$^{1,2}$, V. G. Ksenzov$^2$, A. E. Kudryavtsev$^2$\\
$^{1}$\it\small Department of Mathematics, National Research Nuclear University MEPhI,\\
\it\small Moscow 115409, Russia.\\
$^{2}$\it\small State Scientific Center Institute for Theoretical and Experimental Physics,\\
\it\small Moscow 117218, Russia.\\
}

\date{}

\maketitle

\begin{abstract}

We discuss the case when a fermion occupies an excited non-zero frequency level
in the field of domain wall. We demonstrate that a solution exists for
the coupling constant in the limited interval $1<g<g_{max}\approx 1.65$. We show that
indeed there are different branches of stable solution for $g$ in this interval.
The first one corresponds to a fermion located on the
domain wall ($1<g<\sqrt[4]{2\pi}$). The second branch, which belongs to
the interval $\sqrt[4]{2\pi}\le g\le g_{max}$, describes a polarized fermion
off the domain wall. The third branch with $1<g<g_{max}$ describes an excited
antifermion in the field of the domain wall.

\end{abstract}

\section{Introduction}

In our previous paper \cite{GKK} we studied the problem ''domain wall + excited fermion''.
Initially the problem of the spectrum of a fermion coupled to the field of a static kink
was discussed in Refs.~\cite{DHN, Vol, JR, GW, Stojk}. These papers were
devoted mainly to a zero-frequency fermion bound by the domain wall.

Fermionic bound states in the field of external kink were studied in Ref.~\cite{Vachas}
for the case of the $\lambda\phi^4$-model and in Ref.~\cite{Brih} for the case of
the sine-Gordon model. However, the authors of these publications have considered
kink as given external field. As we shall demonstrate in this our work,
the presence of the fermion changes drastically the kink profile.
So the excitation spectrum for the problem ''fermion coupled to kink''
looks quite different from that calculated in the external field approximation.

We studied the system of the interacting scalar ($\phi$) and fermion ($\Psi$)
fields in two-dimensional space-time $(1+1)$. In terms of dimensionless fields,
coupling constant $g$ and space-time variables $(x,t)$, the Lagrangian density
was taken in the form
\begin{equation}
\mathcal{L}=\frac{1}{2}\left(\partial_\mu\phi\right)^2-\frac{1}{2}\left(\phi^2-1\right)^2
+\bar{\Psi}i\hat{\partial}\Psi-g\bar{\Psi}\Psi\phi.
\label{lagrangian}
\end{equation}
The equation of motion for scalar field $\phi(x,t)$ in the presence of a fermionic
field $\Psi$ reads:
\begin{equation}
\partial_\mu\partial^\mu\phi-2\phi+2\phi^3=-g\bar{\Psi}\Psi,
\label{eqmo}
\end{equation}
where $\bar{\Psi}=\Psi^{\dag}\beta$, $\beta$ is the Pauli matrix, see (\ref{Pauli}).

If the coupling of the scalar field to fermions is switched off, $g=0$, the equation of
motion (\ref{eqmo}) has a static solution called ''kink'',
\begin{equation}
\phi_K(x)=\tanh x.
\label{kink}
\end{equation}
In three space dimensions this solution corresponds to a domain wall that separates
two space regions with different vacua $\phi_{\pm}=\pm 1$, see, e.g. \cite{BelK,Rub} for more
details.

Let us discuss the fermionic sector of the theory. After the substitution
$\Psi(x,t)=e^{-i\varepsilon t}\psi_{\varepsilon}(x)$ the Dirac equation for the
massless case reads:
\begin{equation}
\left(\varepsilon+i\alpha_x\frac{\partial}{\partial x}-g\beta\phi(x)\right)\psi_\varepsilon(x)=0,
\label{eq_psi}
\end{equation}
where $\alpha_x$ and $\beta$ are the Pauli matrices,
\begin{equation}
\alpha_x=
\left(
\begin{array}{lr}
0 & -i\\
i & 0
\end{array}
\right),\quad
\beta=
\left(
\begin{array}{lr}
0 & 1\\
1 & 0
\end{array}
\right).
\label{Pauli}
\end{equation}
In equation (\ref{eq_psi}),
$$
\psi_\varepsilon(x)=
\left(
\begin{array}{l}
u_\varepsilon(x)\\
v_\varepsilon(x)
\end{array}
\right)
$$
is the two-component spinor wave function.
In terms of functions $u_\varepsilon(x)$ and $v_\varepsilon(x)$, Eq.~(\ref{eq_psi}) takes the form
\begin{equation}
\left\{
\begin{array}{l}
\displaystyle\frac{du_\varepsilon}{dx}+g\phi(x)u_\varepsilon=\varepsilon v_\varepsilon,\\
\\
-\displaystyle\frac{dv_\varepsilon}{dx}+g\phi(x)v_\varepsilon=\varepsilon u_\varepsilon.
\end{array}
\right.
\label{system1}
\end{equation}
Substituting $\phi(x)=\phi_K(x)=\tanh x$ we finally get:
\begin{equation}
\left\{
\begin{array}{l}
-\displaystyle\frac{d^2u_\varepsilon}{dx^2}-\frac{g(g+1)}{\cosh^2 x}u_\varepsilon=(\varepsilon^2-g^2) u_\varepsilon,\\
\\
-\displaystyle\frac{d^2v_\varepsilon}{dx^2}-\frac{g(g-1)}{\cosh^2 x}v_\varepsilon=(\varepsilon^2-g^2) v_\varepsilon.
\end{array}
\right.
\label{system3}
\end{equation}
This system describes the spectrum and eigenfunctions of the fermion in the external
scalar field $\phi_K(x)=\tanh x$. Solutions of Eq.~(\ref{system3}) with
$\varepsilon^2\ge g^2$ belong to the continuum and those with $\varepsilon^2<g^2$
to the bound states. The best known discrete mode is the so-called zero-mode
solution (it is time-independent, $\varepsilon=0$):
\begin{equation}
\Psi_{\varepsilon=0}(x)=\sqrt{\frac{\Gamma(g+1/2)}{\sqrt{\pi}\Gamma(g)}}
\left(
\begin{array}{c}
\displaystyle\frac{1}{\cosh^g x}\\
\\
0
\end{array}
\right).
\label{zero_mode}
\end{equation}
For zero-mode solution (\ref{zero_mode}) the r.h.s.\ of Eq.~(\ref{eqmo})
$\bar{\Psi}\Psi=2uv\equiv 0$, so we conclude that the solution of the full problem
in the form ''$\phi_K (x)$ + zero-mode bound fermion (\ref{zero_mode})'' is
self-consistent.

However, if the fermion occupies a level with $\varepsilon\neq 0$, the r.h.s.\ of
Eq.~(\ref{eqmo}) is different from zero. Hence the kink's profile has to be
modified to fulfil Eq.~(\ref{eqmo}). In our previous paper \cite{GKK} we found one
example of analytic solution for the excited fermion on a distorted domain wall, which
indeed is self-consistent.

The plan of this our paper is the following. In Section 2 we study solutions for the first
excited mode. We develop a simple variational procedure that allows to get a
reasonable approximation to the solution for coupling constant $g$ in the interval
$1<g\le g_{max}\approx 1.65$. For $g_1=2\sqrt{2(2-\sqrt{3})}\approx 1.46$
the approximation coincides with the exact solution, found earlier in \cite{GKK}.
In Section 3 we present detailed analysis for the obtained solutions.
In particular, we demonstrate that, depending on the coupling constant, we get
two solution branches for the excited fermion on domain wall. For the coupling
constant in the interval $1<g<\sqrt[4]{2\pi}$ the excited fermion is practically
localized on a distorted domain wall. In the interval of couplings
$\sqrt[4]{2\pi}\le g\le g_{max}$ the initial profile of the domain wall, Eq.~(\ref{kink}),
is almost restored. We also found solution for an excited antifermion (i.e. solution with $\varepsilon<0$)
on the domain wall. This solution looks like a smooth function without cuts
or jumps for any coupling constants $g$ in the interval $g\in(1;g_{max}]$.

At the same time for interval of the coupling $\sqrt[4]{2}<g<g_{max}$
the two components of fermionic wave
function behave very differently with respect to localization.
Namely, the upper component $u_{\varepsilon}(x)$ of the spinor
$\psi_{\varepsilon}(x)$ is located outside the domain wall, while the lower
component $v_{\varepsilon}(x)$ of the fermion wave function sits on the
domain wall. So in this interval of $g$ the domain wall separates in space
the localizations of the upper and the lower components of the spinor.

A general discussion and the summary of the results are presented in Section 4.

\section{Self-consistent solutions for fermion in the field of kink}
\setcounter{equation}{0}

Let us look for a solution of the Dirac equation for the fermion in the field of a distorted
kink $\tilde{\phi}_K=\tanh\alpha x$, where $\alpha$ is unknown real parameter
to be determined from a self-consistency condition, as it will be discussed below.
Introducing a new variable
$y=\alpha x$ and a new parameter $s=g/\alpha$, we get a system of equations
for the fermionic wave function, which formally coincides with that given by Eq.~(\ref{system3}):
\begin{equation}
\left\{
\begin{array}{l}
-\displaystyle\frac{d^2u_{\varepsilon^\prime}}{dy^2}-\frac{s(s+1)}{\cosh^2 y}u_{\varepsilon^\prime}=({\varepsilon^\prime}^2-s^2) u_{\varepsilon^\prime},\\
\\
-\displaystyle\frac{d^2v_{\varepsilon^\prime}}{dy^2}-\frac{s(s-1)}{\cosh^2 y}v_{\varepsilon^\prime}=({\varepsilon^\prime}^2-s^2) v_{\varepsilon^\prime}.
\end{array}
\right.
\label{system4}
\end{equation}
Here $\varepsilon^\prime=\varepsilon/\alpha$. The wave function of the fermion for the
first excited state in the field of the distorted kink $\tilde{\phi}_K$ reads ($1<s<+\infty$):
\begin{equation}
\psi_{\varepsilon^\prime}(x)=
\left(
\begin{array}{l}
u_{\varepsilon^\prime}(x)\\
v_{\varepsilon^\prime}(x)
\end{array}
\right)=
\sqrt{\frac{\alpha\:\Gamma(s-1/2)}{2\sqrt{\pi}\:\Gamma(s-1)}}
\left(
\begin{array}{c}
\displaystyle\sqrt{2s-1}\ \frac{\tanh\alpha x}{\cosh^{s-1}\alpha x}\\
\\
\displaystyle\frac{1}{\cosh^{s-1}\alpha x}
\end{array}
\right),
\label{psi_arb_s}
\end{equation}
Substituting wave function
(\ref{psi_arb_s}) and $\phi_K(x)=\tanh\alpha x$ into equation (\ref{eqmo}), we get
that for $s=2$ and $\varepsilon^{\prime}=\sqrt{3}$ the self-consistency equation
(\ref{eqmo}) is fulfilled if only the slope $\alpha$ satisfies the condition \cite{GKK}:
\begin{equation}
2\alpha^2-2=-\sqrt{3}\alpha^2 \quad \Longrightarrow \quad \alpha^2=2(2-\sqrt{3}).
\label{alpha_cond}
\end{equation}
So for the special case of $s=2$ we obtain an analytic self-consistent solution for the problem
of an excited fermion on the domain wall\footnote{For $s=2$ and $\varepsilon^\prime=-\sqrt{3}$
we get analytic solution for an antifermion on the domain wall. Its wave function is also
given by Eq.~(\ref{psi_arb_s}).}.
This solution corresponds to the value of coupling constant
$$
g=g_1=2\sqrt{2(2-\sqrt{3})}\approx 1.46.
$$

For arbitrary $s\in (1;+\infty)$, $s\neq 2$ there is no analytic solution of equation (\ref{eqmo})
in the form $\phi(x)=\tanh\alpha x$. Substituting (\ref{psi_arb_s}) into (\ref{eqmo})
we obtain the following constraint:
\begin{equation}
2\alpha^2-2=-\frac{\alpha^2 s\:\Gamma(s-1/2)\sqrt{2s-1}}{\sqrt{\pi}\:\Gamma(s-1)}\frac{1}{\cosh^{2s-4}\alpha x}.
\label{eq_arb_s}
\end{equation}
This constraint becomes an algebraic equation for $\alpha$ in the limit $s\to 1$,
(and $g=1$) with $\alpha=1$, and also for $s=2$ (and $g=g_1$) with
$\alpha=\alpha(2)=\sqrt{2(2-\sqrt{3})}$. The latter case corresponds to the exact
solution discussed above, see Eq.~(\ref{alpha_cond}).

However, Eq.~(\ref{eq_arb_s}) may be used to determine the slope $\alpha(s)$ at $x=0$
for the fermionic wave function at arbitrary $s\in (1;+\infty)$.
The function $\alpha(s)$ is shown in Fig.~1. Note that for large $s\gg 1$
$$
\alpha(s)\approx\frac{\sqrt[4]{2\pi}}{s}
$$
and $\displaystyle\lim_{s\to+\infty}\alpha(s)=0$.

\begin{figure}[t]
\hspace{5mm}
\includegraphics[width=9cm, height=15cm, angle=-90]{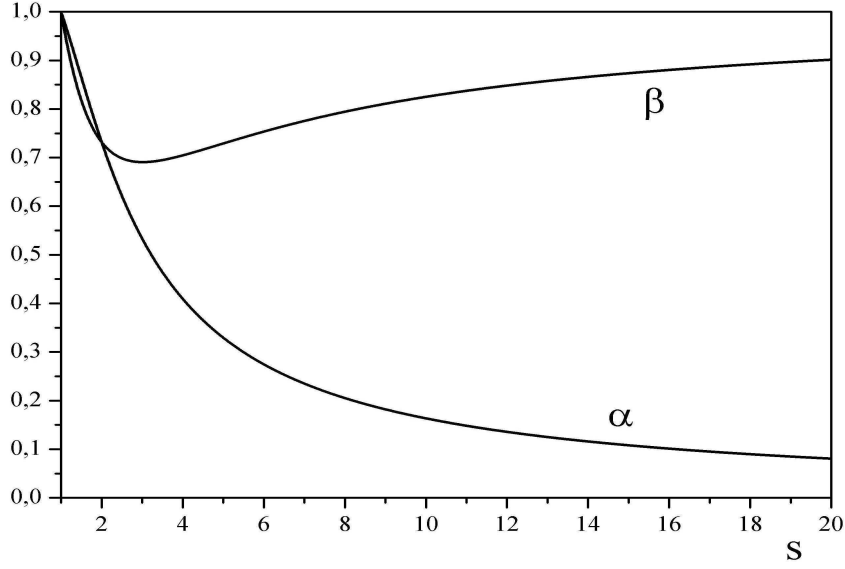}
\label{fig1}
\caption{slopes $\alpha(s)$ and $\beta(s)$.}
\end{figure}

Inserting fermionic wave function (\ref{psi_arb_s}) with the so obtained $\alpha(s)$
into the r.h.s.\ of Eq.~(\ref{eqmo}), we get the following equation for the scalar field $\phi(x)$:
\begin{equation}
\frac{d^2\phi}{dx^2}+2\phi-2\phi^3=\frac{\alpha^2(s)\:s\:\sqrt{2s-1}\:\Gamma(s-1/2)}{2\sqrt{\pi}\:\Gamma(s-1)}
\frac{\tanh(\alpha(s)\:x)}{\cosh^{2s-2}(\alpha(s)\:x)}.
\label{eq_phi}
\end{equation}
This equation has no solutions for boundary conditions
$\phi_s(0)=0$, $\displaystyle\frac{d\phi_s(0)}{dx}=\alpha(s)$, $\phi\to 1$ with $x\to +\infty$ at
arbitrary $s\in (1;+\infty)$. The only exceptional cases are those with $s=1$ and
$s=2$, for which we found exact solutions discussed above. However, one may
try to solve the equation of motion (\ref{eq_phi}) with modified boundary conditions
for the field $\phi_s(x,0)$:
\begin{equation}
\phi_s(0)=0, \quad \frac{d\phi_s(0)}{dx}=\beta(s), \quad \lim_{x\to+\infty}\phi_s(x)=1.
\label{bound_cond}
\end{equation}
We solved Eq.~(\ref{eq_phi}) with boundary conditions (\ref{bound_cond})
numerically, using the shooting method for the whole interval of
parameter $s\in(1;+\infty)$. The resulting function $\beta(s)$ is shown in Fig.~1. Note
that $\beta(s)=\alpha(s)$ at two points $s=1$ and $s=2$, for which our procedure
reproduces exact solutions. In the region of small $s\in(1;2)$ functions $\beta(s)$
and $\alpha(s)$ do not differ drastically.
On the other hand, $\displaystyle\lim_{s\to+\infty}\beta(s)=1$, what means that our solution
for scalar field approaches undistorted kink in the limit of large $s$.

\begin{figure}[t]
\includegraphics[width=5cm, height=8cm, angle=-90]{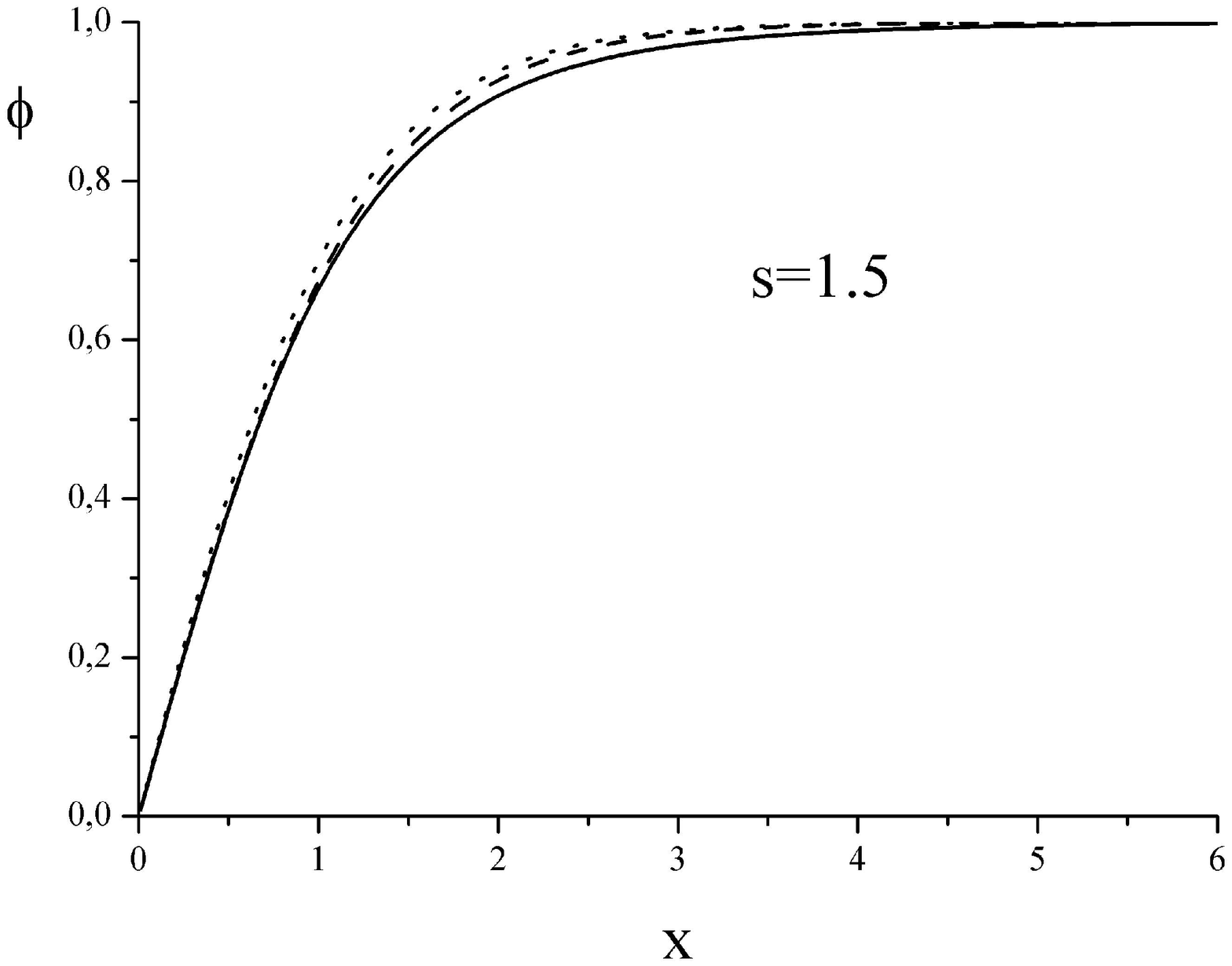}
\includegraphics[width=5cm, height=8cm, angle=-90]{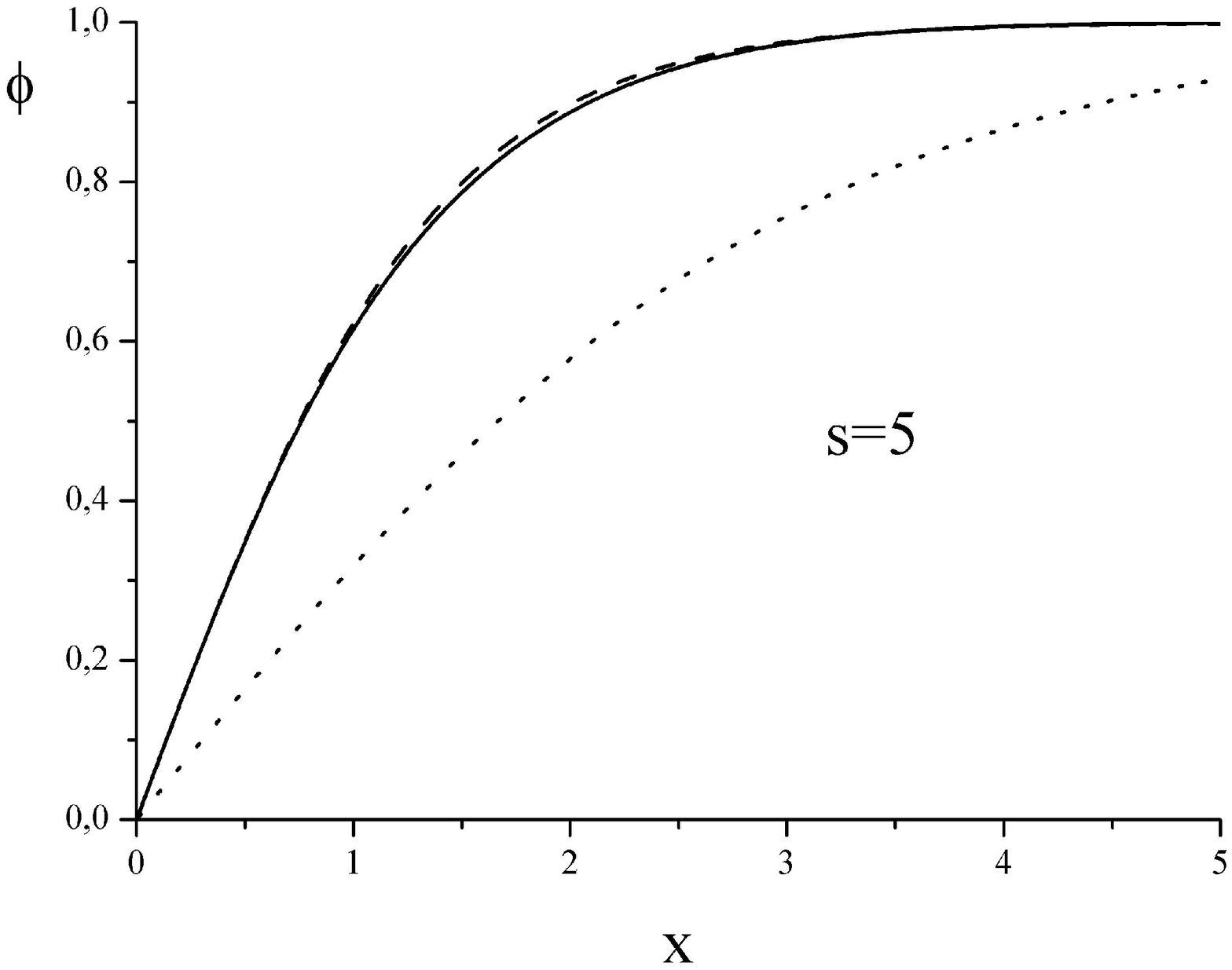}

\hspace{40mm}
\includegraphics[width=5cm, height=8cm, angle=-90]{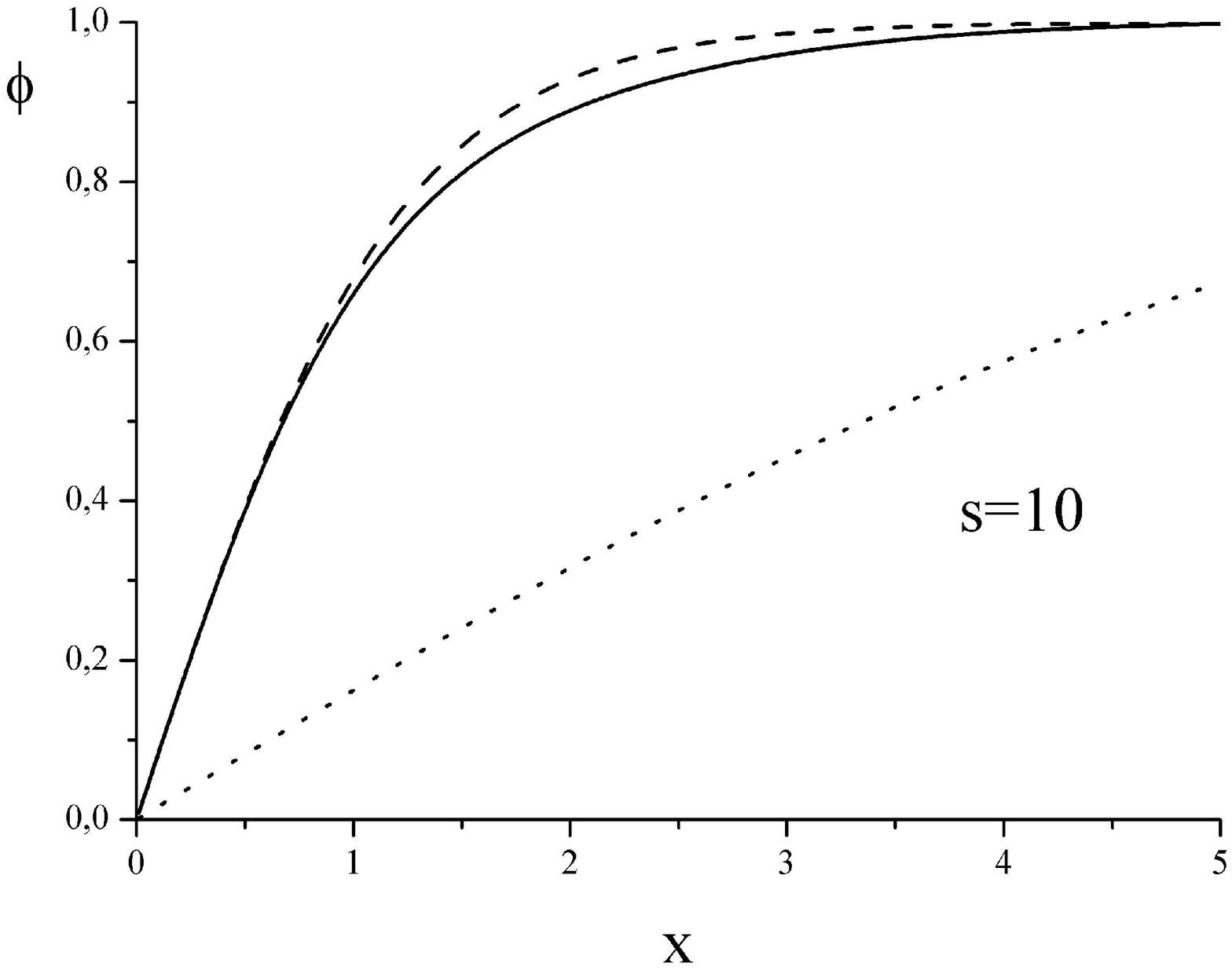}
%
%
\label{fig2}
\caption{profiles of scalar field $\phi_s(x)$ at different values of parameter $s$.
Solid line -- numerical solution of eq.~(\ref{eq_phi}); dashed line -- approximation of the exact
solution by $\phi_s(x)=\tanh(\beta(s)x)$. Dotted line -- the function $\phi_s^{in}(x)=\tanh(\alpha(s)x)$.}
\end{figure}

Numerical solutions for the scalar field $\phi_s(x)$ at different values
of parameter $s$ are given in Fig.~2, where we also show the profile of the function
$\tilde{\phi}_s(x)=\tanh(\beta(s)x)$.
This figure illustrates that for all  $s\in(1;+\infty)$ the exact numerical solution is very
close to $\tanh(\beta(s)x)$.
So one can say that with high accuracy the field configuration $\phi_s(x)$ is
a distorted kink with the slope $\beta(s)$ at the origin.

In terms of fields $\{\phi_s(x);\Psi_s(x)\}$ the procedure we used here is indeed
a variational-type one. This approximate procedure reproduces exact solutions
for $s=1$, 2 and gives reasonable results in the whole range of parameter $s$.

\section{Stable branches of solution for a fermion coupled to domain wall}

\setcounter{equation}{0}

The energy of a massless fermion in the field of the distorted kink $\phi_K(\alpha(s)x)$
is $\varepsilon(s)=\sqrt{2s-1}\ \alpha(s)$ and $-\varepsilon(s)$ for massless antifermion.
The function $\varepsilon(s)$ is shown in
Fig.~3a. Note that for large $s\gg 1$ $\varepsilon(s)$ goes to zero,
$\varepsilon(s)\approx\displaystyle\frac{\sqrt[4]{8\pi}}{\sqrt{s}}$.

\begin{figure}[t]
\includegraphics[height=8cm, width=6cm, angle=-90]{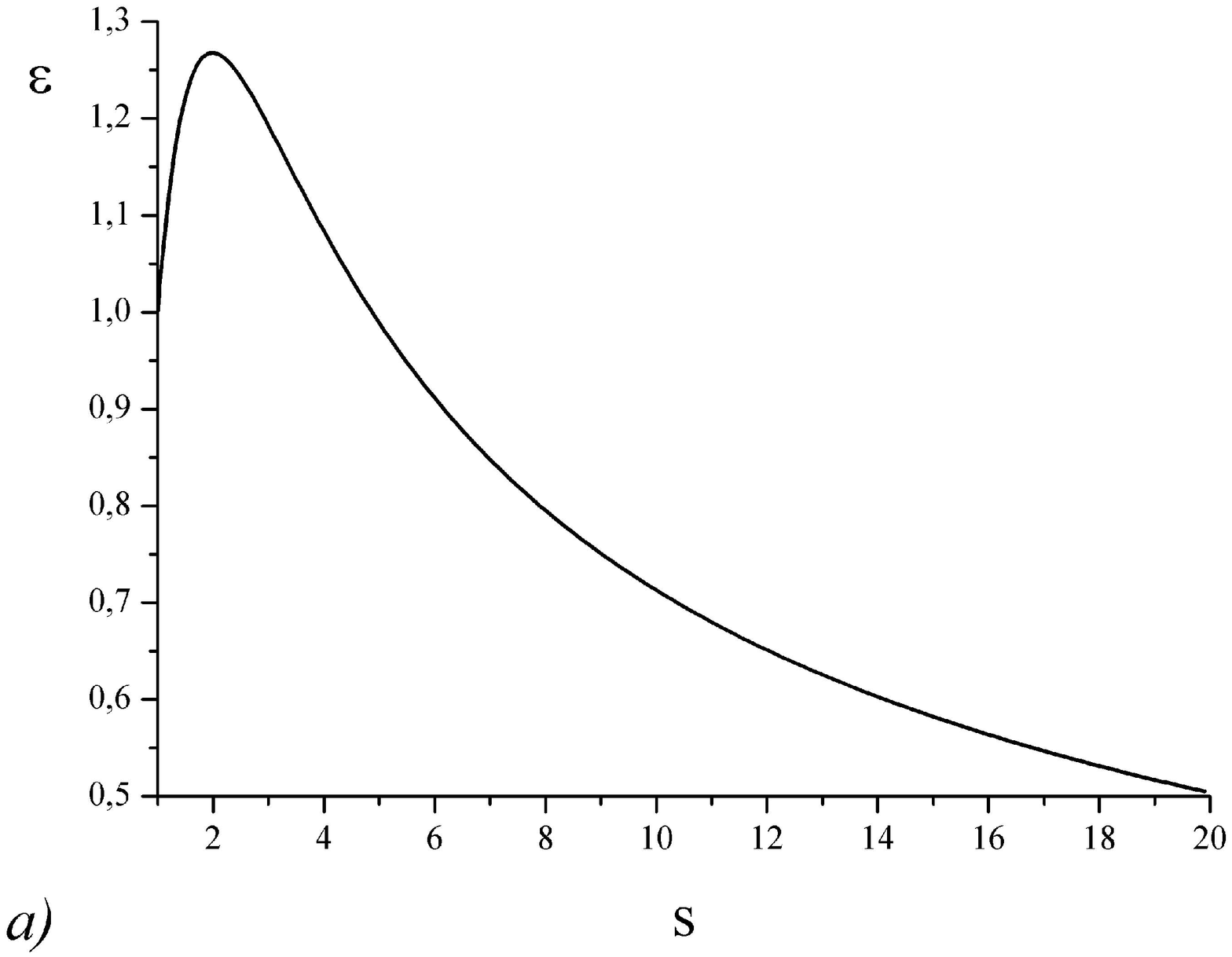}
\includegraphics[height=8cm, width=6cm, angle=-90]{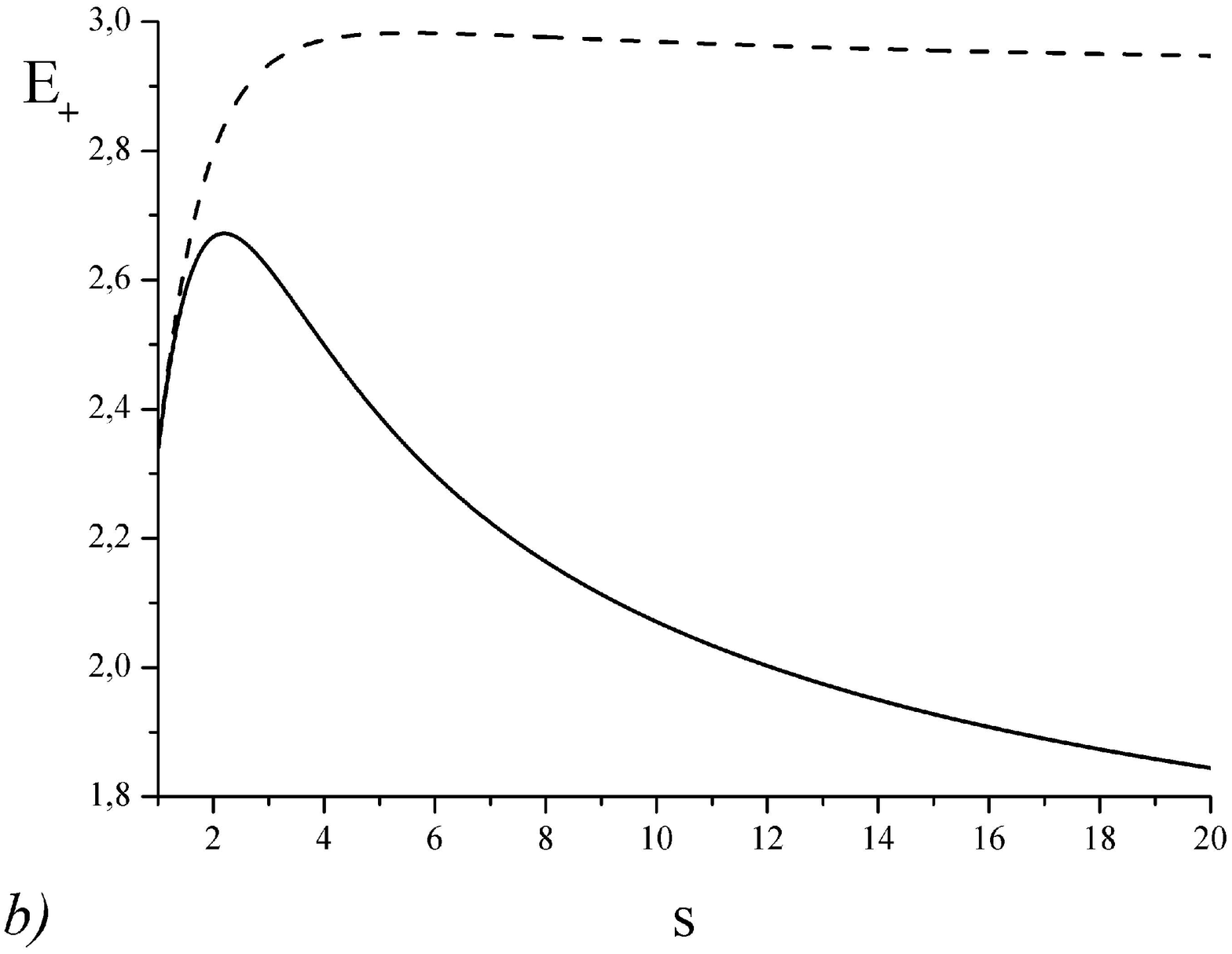}
\label{fig3}
\caption{a) the energy of the fermion field $\varepsilon(s)$; b) total energy of the system
''kink + excited bound fermion'' $E_{+}(s)$ (solid line), the boundary line of the region
''domain wall + fermion in continuum'' (dashed line).}
\end{figure}

The energy of the system ''kink + bound fermion (antifermion)'' may be well approximated by
the following simple expression:
\begin{equation}
E_{\pm}(s)=\frac{2}{3}\left(\beta(s)+\frac{1}{\beta(s)}\right)\pm\varepsilon(s).
\label{energy}
\end{equation}
The function $E_{+}(s)$ is shown in Fig.~3b. Notice that for each $s\in (1;+\infty)$
the energy of the field configuration $E_{+}(s)$ is smaller than the energy of
the configuration ''undistorted domain wall + fermion at rest in continuum''. The energy of the latter
configuration $E_{free}(s)=\displaystyle\frac{4}{3}+g(s)$ is also shown in Fig.~3b.
Looking at this figure we conclude that the field configuration ''kink + excited fermion'' is truly bound.

Now let us consider the dependences of observables in terms of coupling
constant $g=g(s)=s\:\alpha(s)$. The curve $g(s)$ is shown in Fig.~4. As it is seen,
the function $g(s)$ reaches its maximum $g_{max}\approx 1.65$ at
$s_m\approx 5.58$ and falls down for large $s\gg 1$ reaching
$g_{as}=\sqrt[4]{2\pi}\approx 1.58$ at $s\to+\infty$. The asymptotic behaviour
of $g(s)$ is:
$$
g(s)=\sqrt[4]{2\pi}\left(1+\frac{3}{8s}\right).
$$
The crossing point $g(s_1)=g_{as}$ is also marked in Fig.~4 with $s_1\approx 2.78$.

\begin{figure}[t]
\hspace{5mm}
\includegraphics[height=15cm, width=9cm, angle=-90]{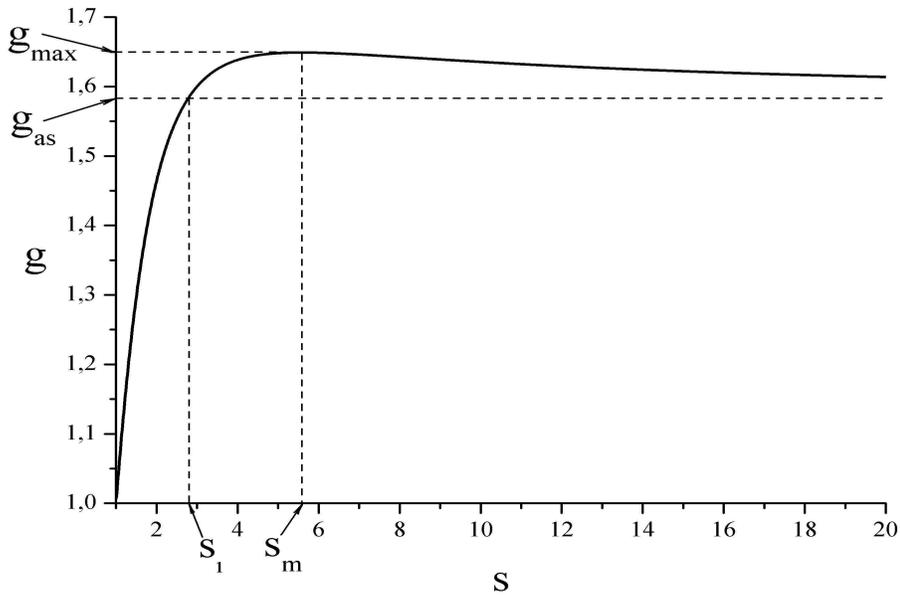}
\label{fig4}
\caption{coupling constant $g$ depending on parameter $s$.}
\end{figure}

The energy of the system $E_{\pm}(g)$ as a function of coupling constant
is shown in Fig.~5. Note that in the
interval $1\le g<g_{as}$ ($1\le s<s_1$) the functions $E_{+}(g)$ and $E_{-}(g)$ are single-valued, whereas
in the interval $g_{as}<g\le g_{max}$ ($s_1\le s<+\infty$) they are
double-valued. So for each $g\in (g_{as};g_{max}]$ there are different branches of the solution
for both $E_{+}(s)$ and $E_{-}(s)$.
The first branch of function $E_{+}(s)$ corresponds to the interval $s\in [s_1;s_m]$. The second
branch corresponds to $s\in [s_m;+\infty)$. As it is seen from Fig.~5, the
first branch of solution $E_{+}(s)$ has larger energy $E(g(s))$ than the second one. That is
why the first branch is unstable with respect to the decay into the
second branch. The excess of energy could be emitted in the form of waves of scalar field
$\phi$. The unstable branch of the solution is drawn in Fig.~5 by dotted line.
Analogous situation takes place for function $E_{-}(s)$. Stable branch of solution for
function $E_{-}(s)$ is also shown in Fig.~5 by solid line.

\begin{figure}[t]
\hspace{5mm}
\includegraphics[height=15cm, width=9cm, angle=-90]{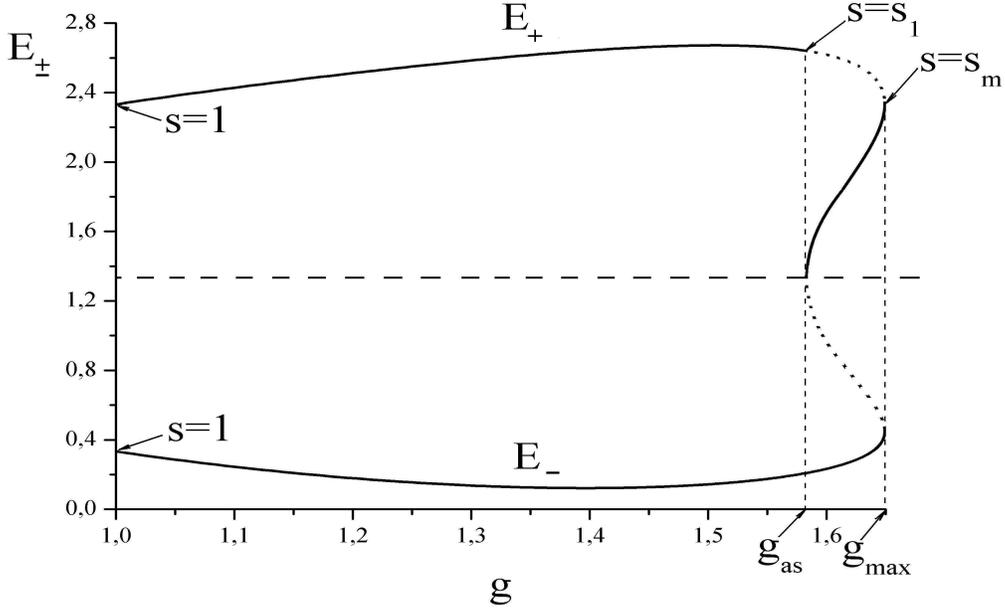}
\label{fig5}
\caption{total energy $E_{\pm}$ of the system as a function of coupling constant $g$;
stable branches of the solution are shown by solid lines; unstable branches are
indicated by dotted lines. Horizontal dashed line demonstrates the energy of
the udistorted kink.}
\end{figure}

The lower stable branch for function $E_{+}(s)$ corresponds to large $s\in [s_m;+\infty)$.
As it follows from Fig.~3a, in this region $\varepsilon(s)$ is small and
$\varepsilon(s)\to 0$ at $s\to+\infty$ (or $g\to g_{as}$). That is why it is quite natural
to call this branch the ''quasi zero-mode'' solution.

So we got two distinct stable branches for field configurations: domain wall carrying
an excited fermion. Let us discuss what these field configurations look like.

\begin{figure}[t]
\hspace{5mm}
\includegraphics[height=15cm, width=9cm, angle=-90]{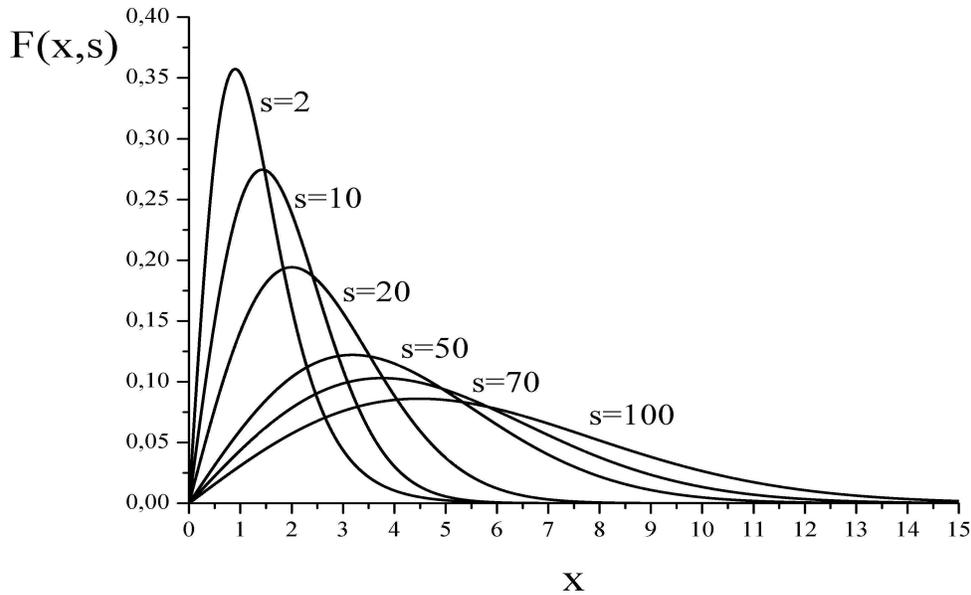}
\label{fig6}
\caption{the right-hand side of eq.~(\ref{eq_phi}) as a function of $x$ for some values of $s$.}
\end{figure}

The r.h.s.\ of Eq.~(\ref{eq_phi}) $F(x,s)$ is proportional to the product
$u_{\varepsilon^\prime}(x)v_{\varepsilon^\prime}(x)$. For different values of parameter
$s$ the function $F(x,s)$ is shown in Fig.~6. The maximum position for each curve
is at $x_{max}=\displaystyle\frac{\sqrt{s}}{\sqrt[4]{8\pi}}$ ($s\gg 1$), and
$$
F(x_{max},s)=\sqrt{\frac{2}{es}}\ .
$$
So in the limit of large $s$, i.e.\ for $g\to g_{as}=\sqrt[4]{2\pi}$, the r.h.s.\ of Eq.~(\ref{eq_phi})
becomes small and as a result the profile of the domain wall looks similar to the unperturbed
kink $\phi_K=\tanh x$.

Let us clarify what the wave function of fermion looks like in the limit of large $s\gg 1$.
The position of the maximum of $u^2_{\varepsilon^\prime}(x)$ is at
$\tilde{x}_{max}=\displaystyle\frac{\sqrt{s}}{\sqrt[4]{2\pi}}$ and
$$
u^2_{\varepsilon^\prime}(\tilde{x}_{max})=\sqrt[4]{\frac{2}{\pi e^4}}\ \frac{1}{\sqrt{s}}\ .
$$
Some profiles of $u^2_{\varepsilon^\prime}(x)$ are
presented in Fig.~7a. We conclude that in the limit of large $s$ the upper
component of the fermionic wave function is located outside the domain wall.

\begin{figure}[t]
\includegraphics[height=8cm, width=6cm, angle=-90]{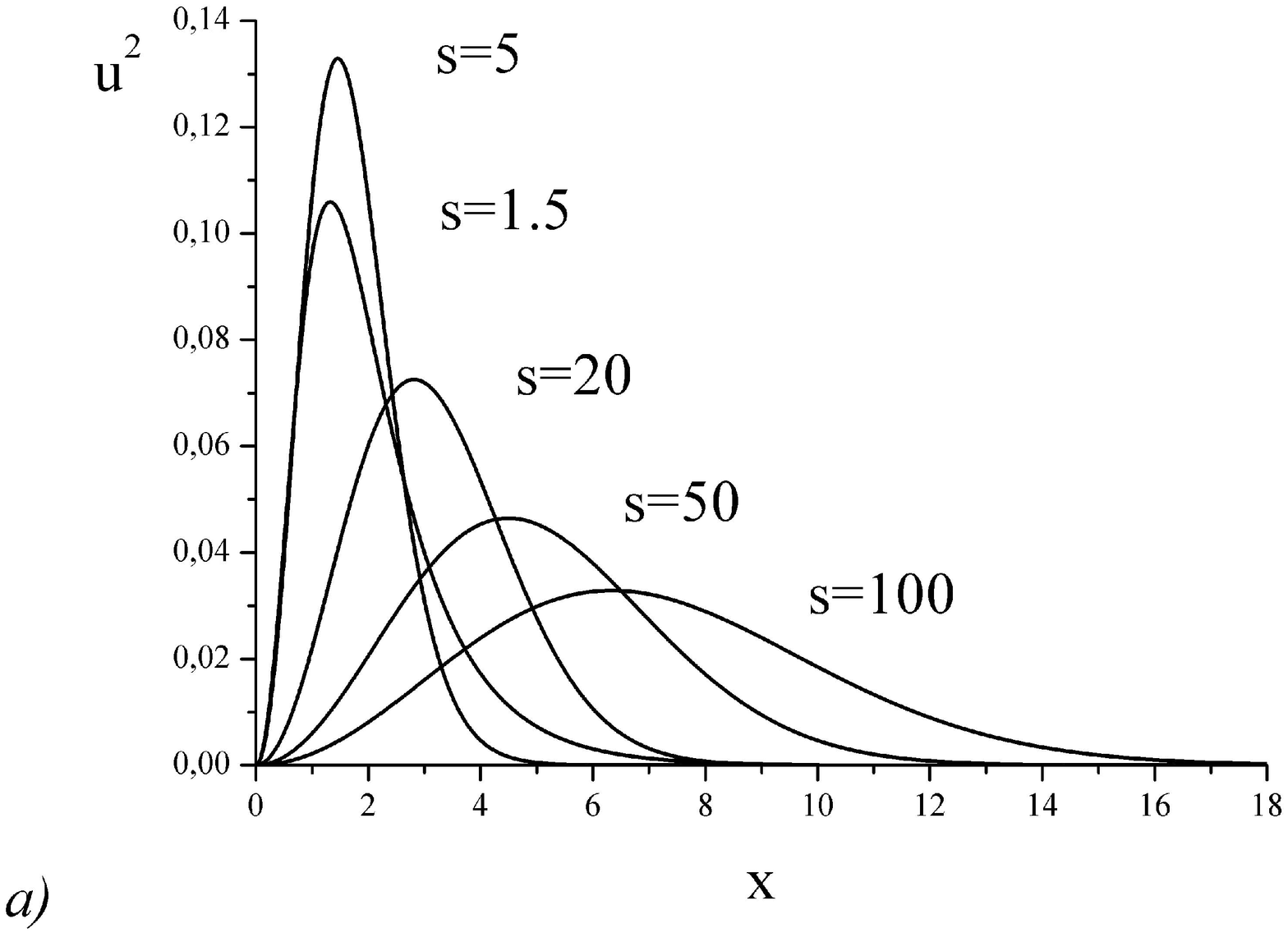}
\includegraphics[height=8cm, width=6cm, angle=-90]{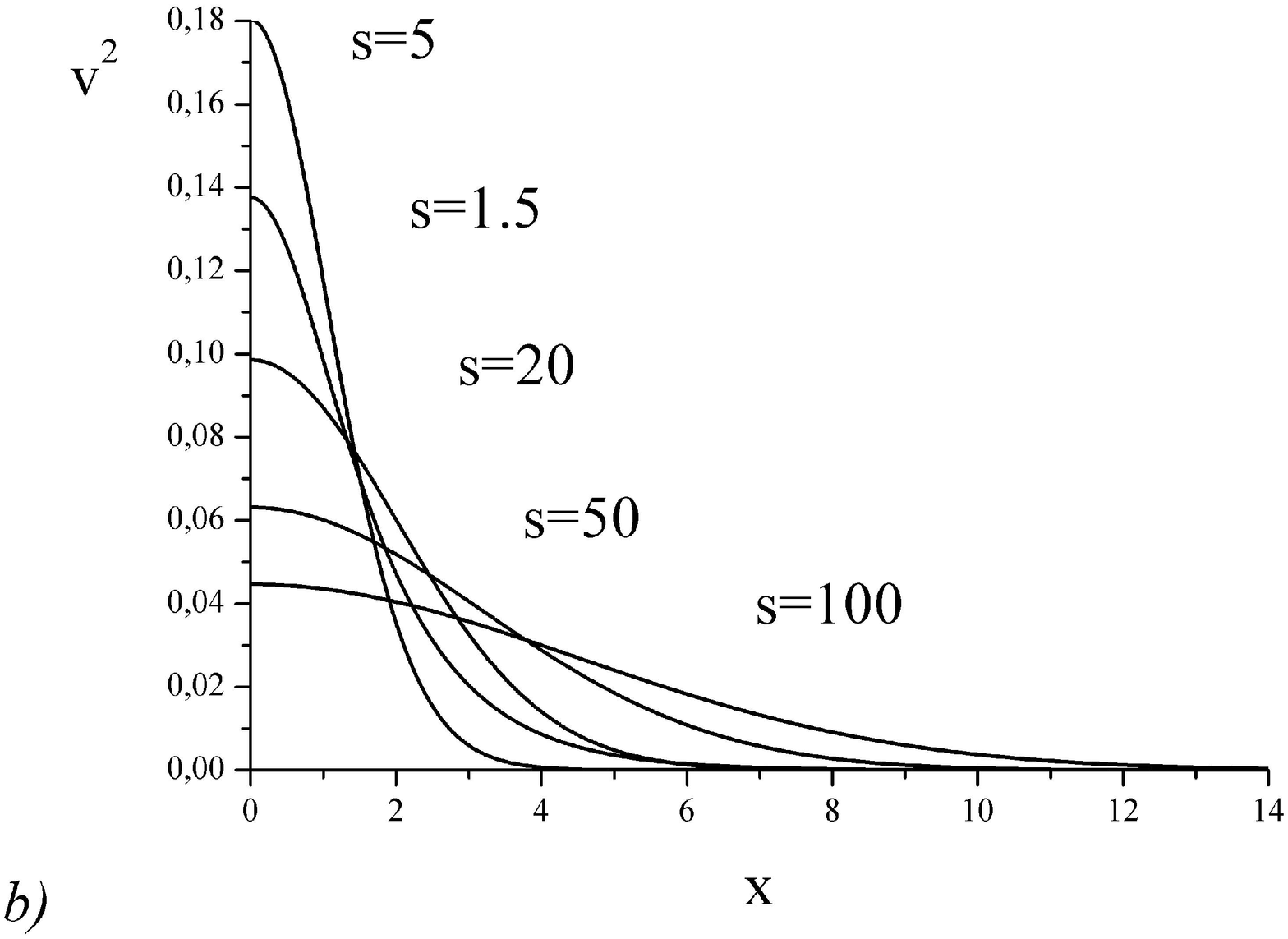}
\label{fig7}
\caption{the functions $u^2_{\varepsilon^\prime}(x)$ (figure a) and
$v^2_{\varepsilon^\prime}(x)$ (figure b) for some values of $s$.}
\end{figure}

On the contrary, the maximum of the lower component $v_{\varepsilon^\prime}(x)$
is at $x=0$, i.e.\ on the domain wall, for all values of $s\in (1;+\infty)$ and
$$
v^2_{\varepsilon^\prime}(0)=\frac{1}{\sqrt[4]{8\pi}}\frac{1}{\sqrt{s}}\ .
$$
In the limit of large $s$,
$v_{\varepsilon^\prime}(x)\sim\exp(-g_{as}|x|)$ ($|x|\to+\infty$).
The function $v_{\varepsilon^\prime}(x)$ behaves as a zero-mode solution in the
limit of large $s$. Some profiles of $v^2_{\varepsilon^\prime}(x)$ are presented in Fig.~7b.

Thus, in the limit of large $s$ the wave function of a fermion in the first excited state practically
splits into two parts. The upper component of the wave function,
$u_{\varepsilon^\prime}(x)$, which describes fermions with spin projection $+1/2$,
is located outside the domain wall. The lower component of the wave function,
$v_{\varepsilon^\prime}(x)$, with spin projection $-1/2$, prefers smaller distances from domain wall.
As a result, in the limit of large $s$ the domain wall separates the maxima of the upper
$u_{\varepsilon^\prime}(x)$ and the lower $v_{\varepsilon^\prime}(x)$ components
of the fermionic wave function in space.

\section{Conclusion}

\setcounter{equation}{0}

We used a simple variational method and found a solution of the problem
''domain wall + excited fermion''. We performed numerical simulations.
For each coupling constant $g\in (1;g_{max}]$ the solution is a bound fermion with
wave function (\ref{psi_arb_s}) in the field of the distorted kink $\tilde{\phi}_K=\tanh(\beta(s)x)$.
The functions $\alpha(s)$ and $\beta(s)$ are shown in Fig.~1. Parameter $s$ varies
in the interval $1<s<+\infty$. This corresponds to a variation of coupling constant $g$
in the limits
$$
1<g\le g_{max}\approx 1.65.
$$
No solution for a fermion in the first excited state on the domain wall was found
for any $g>g_{max}$.

For $g\in (1;g_{max}]$ we found two separate stable solution branches
for function $E_{+}(s)$. For $s\in (1;s_1)$, where $s_1=2.78$ and
$g(s_1)=g_{as}=\sqrt[4]{2\pi}\approx 1.58$, the fermion wave function
is located on the distorted domain wall. For $s=2$ ($g=g_1=2\sqrt{2(2-\sqrt{3})}$),
belonging to this interval, our variational procedure reproduces the exact self-consistent
solution of the problem, obtained in our previous publication \cite{GKK}.

For $s\in [s_1;s_{m}]$, where $s_{m}=5.58$, the solution is unstable.

For the interval $s\in [s_{m};+\infty)$, which corresponds to the variation of coupling
constant $g\in [g_{max};g_{as})$, the fermionic wave function consists of two different
pieces approximately separated in space. The upper component
$u_{\varepsilon^\prime}(x)$ is located outside the domain wall and the lower one
$v_{\varepsilon^\prime}(x)$ prefers smaller distances. Note that
$\displaystyle\int\limits_{-\infty}^{+\infty}|v_{\varepsilon^\prime}(x)|^2dx=\frac{1}{2}$,
which means that only one half of the fermionic wave function is located near the domain wall
in this regime of large $s$. The profile of the domain wall for these large
$s$, almost restores its original profile $\phi_K=\tanh x$, as $\beta(s)\to 1$
at $s\to +\infty$.

Note that the energy of the fermion, $\varepsilon(s)$, is getting very small in the limit
of large $s$ (see also Fig.~3a). We conclude that in this limit, $s\gg 1$, the spectrum
of the problem ''domain wall + fermion'' has two nearly degenerate states: the
zero-mode ground state ($\varepsilon=0$) and the state with an excited fermion.
The energy of the latter is
$\varepsilon(s)\approx\displaystyle\frac{\sqrt[4]{8\pi}}{\sqrt{s}}\ll 1$, so this level
may be classified as a quasi zero-mode solution. For $g\ge g_{max}$ only zero-mode
solution survives.
The picture is slightly different for an antifermion on the domain wall
(function $E_{-}(s)$). For this case the stable branch of solution corresponds
to $s\in (1;s_m)$ ($g\in (1;g_{max})$).

Speaking of the practical applications of the results discussed above, we should
first of all mention the problems of cosmology. In this connection we refer to the classical
studies of Refs.~\cite{Zel,Vol2,Berezin}.

Note that a similar problem ''kink + charged scalar field'' at classical level was
solved in Ref.~\cite{Lensky}. A related problem of the interaction of a domain wall
with a skyrmion was studied in Ref.~\cite{KPZ}.

Some other properties of fermions in the fields of topological objects such as solitons
and instantons one may also find in the monography by R.~Rajaraman~\cite{Rad}.

\section {Acknowledgments}
The authors are thankful to A.~A.~Abrikosov, O.~V.~Kancheli and W.~J.~Zakrzewski for
useful discussions and to V.~A.~Lensky for his careful reading of the manuscript and for
many useful remarks. This work was partially supported by the Russian State Atomic
Energy Corporation ''Rosatom'' and by grant NSh-4172.2010.2.

\end{document}